\documentstyle[preprint,aps]{revtex}
\begin{document}
\draft

\title{Elastic String in a Random Medium}

\author{Hern\'an A. Makse,  Albert-L\'aszl\'o Barab\'asi$^*$ and H. Eugene
Stanley}

\address{Center for Polymer Studies and  Department of Physics,
Boston University,  Boston, MA  USA}

\date{\today}

\maketitle

\begin{abstract}
We consider a one dimensional elastic string as a set of massless
beads interacting through springs characterized by anisotropic elastic
constants. The string, driven by an external force, moves in a medium
with quenched disorder.  We present evidence that the consideration of
longitudinal fluctuations leads to nonlinear behavior in the equation
of motion which is {\it kinematically} generated by the motion of the string.
The strength of the nonlinear effects depends on the
anisotropy of the medium and the distance from the depinning
transition. On the other hand the consideration of restricted solid on
solid conditions  imposed to the growth of the string leads to a
nonlinear term in the equation of motion with a {\it diverging} coefficient at
the depinning transition.
\end{abstract}
\narrowtext

The motion of an elastic string in disordered media has attracted
considerable recent attention, in part due to its relevance to
flux-flow in type-II superconductors \cite{super} and roughening of
non-equilibrium
interfaces \cite{Vicsek}.  By means of a number of numerical
\cite{Dong93,Kaper93,Leschorn93,lech2,Chang94,makse1} and analytical
\cite{Natter92,Narayan93} studies it has been observed that scaling
theory can be used as an underlying framework to understand and
characterize the dynamical properties of the elastic string.

Consider a one-dimensional elastic string moving under
the influence of an external driving force $F$, normal to the
string, in a two-dimensional
disordered medium of edge $L$.
A discrete model for such a string consists on $N$ massless beads connected
by springs. The string is assumed to be oriented along the $x$-axis
 and the position of the $i$-th bead is denoted by a two-dimensional
displacement vector $\vec{r}_i \equiv (x_i , y_i)$, $i=1,\ldots,L$
(see
Fig. \ref{sketch}).
The disorder in the medium is introduced by uniformly-distributed
pinning sites with random strength, which we refer to as quenched disorder or
``quenched noise''.
The dynamics of such a string is the result of the interplay
 between the quenched disorder characteristic of the medium and the elastic
properties of the string.

A key  quantity is the average
velocity of the string as a function of the external force.
At small  forces $F$ the string is pinned by
static disorder. Just above the depinning
transition $F=F_c$, i.e.
when the external force overcomes the pinning effect of impurities,
the velocity varies as
\begin{equation}
\label{v}
v_{0} (f)  \sim  f^\theta ,
\end{equation}
where $\theta$ is the  velocity exponent and $f \equiv F/F_c-1$
the reduced force.

Neglecting
thermal fluctuations and lateral fluctuations of the beads,
the equation of motion for the string in the
continuum limit  is the Edwards-Wilkinson equation \cite{ew}
with quenched disorder
\cite{Dong93,Kaper93,Leschorn93,lech2,Chang94,makse1,Natter92,Narayan93}
\begin{equation}
\label{lin}
\frac{\partial y(x,t)}{\partial t} = \nu \nabla^2 y + \eta(x,y) + F .
\end{equation}
The first term in the right hand side of (\ref{lin})
includes the elastic effects acting to make the string straight.
The second term mimics the quenched  disorder,
which  has  zero mean and is uncorrelated.
The string is driven in the $y$ direction   by the external force $F$.
For large driving force
$(F \gg F_c)$,
the quenched noise becomes effectively time-dependent,  $\eta(x, y_0 +vt)$.
It is believed \cite{Narayan93,Parisi92} that in this regime
the motion of the string induces
an additional nonlinear term in (\ref{lin}), namely the
Kardar-Parisi-Zhang (KPZ) term
$ \lambda (\nabla y)^2$ \cite{KPZ}.
However, since this nonlinear term is generated by the motion of the
string, $\lambda$ is expected to vanish as the velocity goes to zero at
the depinning transition, and the critical behavior at $F=F_c$ is
correctly described by Eq. (\ref{lin}).

While  (\ref{lin}) can be obtained (using $\partial_t y=
- \delta {\cal H} / \delta y + F $)
from the Hamiltonian
\begin{equation}
\label{hami1}
{\cal H} =  \int^L_0  dx \{ \nu (\nabla y)^2 + \mu(x,y) \},
\end{equation}
the KPZ nonlinear term $ \lambda (\nabla y)^2$ cannot
be deduced as a variation of any bounded Hamiltonian.
Here the quenched noise is $\eta(x,y) = - \delta_y \mu(x,y)$.

In the Hamiltonian (\ref{hami1}), only  transverse fluctuations
(along the $y$ direction) contribute
to the elastic energy, forbidding longitudinal  fluctuations
(along the $x$ direction).
However, for a real elastic string, the elastic
energy depends on the distance $(\vec{r}_i - \vec{r}_{i-1})^2$
between two consecutives beads.
Here we introduce a (1+1)-dimensional model which allows both
for longitudinal
and transverse
fluctuations of the beads.
In the model, the elastic energy depends both on $\nu_x (x_i-x_{i-1})^2$ and
$\nu_y (y_i-y_{i-1})^2$, where $\nu_x$ and $\nu_y$ are the elastic constants
corresponding to displacements in the $x$ and $y$ direction,
respectively. We focus on the determination of the equation of motion
of the string. We find that, even though the string can form overhangs,
at large enough length scales
the string  still have a well-defined orientation and
profile, and can be described by a continuum theory.
The main results of this paper are:

{\bf (a)} In the limit,  $\varepsilon \equiv \nu_y / \nu_x \gg 1$,
where  $\varepsilon$ is the
anisotropy
parameter,
the
large-scale behavior of
the string is described by the nonlinear equation of motion with quenched
noise
\begin{equation}
\label{kpz}
\frac{\partial y(x,t)}{\partial t} = \nu \nabla^2 y +  \lambda (\nabla y)^2 +
 \eta(x,y) + F ,
\end{equation}
where the nonlinear term   $\lambda (\nabla y)^2$ in
(\ref{kpz}) is of kinematic origin.
We find that $\lambda$ vanishes at the depinning
transition as
\begin{equation}
\label{lam-conv}
\lambda(f) \sim f^{|\phi|} \to 0.
\end{equation}

{\bf  (b)} If longitudinal fluctuations are  neglected, nonlinear terms
of the type $ \lambda (\nabla y)^2$ are forbidden in the growth
equation. We argue that this result applies to a number of
previously-introduced models
\cite{Dong93,Kaper93,Leschorn93,lech2,Chang94,makse1}.
In our model, this limit corresponds to
 $\varepsilon \equiv \nu_y / \nu_x \to 0$.

{\bf (c)} A different scenario is found when the rules of motion of the
beads are constraint to satisfy a restricted solid on solid (RSOS)
condition  $|h_{i\pm 1}-h_{i}| \le const$ \cite{kk}.
When such condition is imposed we find that the equation of
motion of the string
 is (\ref{kpz}) but with a coefficient
$\lambda$ which diverges at the depinning
transition as
\begin{equation}
\label{lam-div}
\lambda(f) \sim f^{-\phi} \to \infty.
\end{equation}
This result is valid for any value of the anisotropy
parameter $\varepsilon$, and
applies to a number of growth model in the directed percolation
universality class \cite{lech3,sergey,snepen,amaral,amaral2,makse2}.

We now take up each of these results in turn. Before beginning, we note that
for a given model the presence of a nonlinear term
$\lambda (\nabla y)^2$ can be identified using tilt-dependent velocity
measurements \cite{Krug,amaral,amaral2,makse2}.
Suppose we
tilt  the  elastic string, by imposing
helical boundary conditions $y_1 = y_L + m L$, where $m$ is the
average tilt of the string. Then, according to (\ref{kpz}), the
average tilt-dependent velocity becomes
\begin{equation}
\label{vel}
v(m) = v_0 + \lambda m^2 ,
\end{equation}
where $v_0$ is the  velocity of the untilted string.
If $\lambda=0$,
so that the motion
of the elastic string is described by (\ref{lin}),
 then the  velocity does not depend on the average tilt
of the interface. Tilt dependence is expected
only if there is a  nonlinear term in the equation of motion
of the form $\lambda (\nabla y)^2$.
This property can be used to gain information on the presence and
magnitude of the  nonlinear term $\lambda$,
by monitoring the velocity
of the string  as a function of the average tilt, and fitting to a
parabola the obtained curve \cite{largeslope}.

In the following, we study a generalized model
of  an elastic string that allows for lateral motions of the beads and
therefore overhangs. The main element of the model,
not included in the Hamiltonian
(\ref{hami1}), is the existence of longitudinal motion
of the beads. To include this additional degree of freedom,
we use a generalized Hamiltonian
\begin{equation}
\label{hami2}
\begin{array}{ll}
{\cal H} =   \sum^{L}_{i=1}&
 [ \nu_x (x_i - x_{i-1})^2 + \nu_y (y_i - y_{i-1})^2 + \\
 & \mu(x_i,y_i) - F  y_i ].
\end{array}
\end{equation}

We simulate  the discrete version of (\ref{hami2}),
concentrating on the zero-temperature
dynamics of the string (only motions which decrease the total energy of the
string are allowed).
A standard Monte Carlo
algorithm, by choosing randomly a site on the interface,
induces time-dependent noise.
Since
at zero temperature the motion of the string is deterministic,
we have employed  an algorithm with parallel updating, during which
even and odd sublattices are updated simultaneously.
The quenched noise is introduced by defining at every site of the
two-dimensional lattice uncorrelated
random numbers $\mu(i,j)$, uniformly distributed  between $-\delta$ and
$\delta$.

During the simulations, the chosen bead is allowed to move to one of
its {\it four} nearest neighbors, if that motion decreases the total
energy of the string given by (\ref{hami2}).  If there are more than
one possible moves with $\Delta {\cal H} < 0$, then the one with most
negative $\Delta {\cal H}$ is chosen.  We focus on the determination
of the nonlinear term $\lambda$, measuring the tilt-dependent velocity
of the string.

{\bf (a)} Figure \ref{velocity}a shows the velocity of the driven
elastic string as a function of the average tilt for different driving
forces.  The results correspond to the anisotropic motion
characterized by $\nu_x= 0.1$ and $\nu_y= 1$ $(\varepsilon = 10)$.  We
see that the velocity follows a parabola with the tilt, indicating the
presence of a nonlinear term $\lambda (\nabla y)^2$ above the
depinning transition (moving phase, $F>F_c$). However, the parabolas become
flatter as the depinning transition is approached. Our calculations
indicate that $\lambda \to 0$ as $F \to F_c$ as in (\ref{lam-conv}).
These results are obtained for the anisotropic case $\nu_x < \nu_y$,
($\varepsilon > 1)$, and further increasing the anisotropy, the
observed behavior does not vanish.

{\bf (b)} The other limit of the model leads to known results:
a finite  $\nu_y$ and $\nu_x \to \infty$
means that  longitudinal fluctuations are energetically very expensive,
allowing only transversal fluctuations.
In this limit, the model reduces to the  models of Refs.
\cite{Dong93,Kaper93,Leschorn93,lech2,Chang94,makse1},
where longitudinal fluctuations are  not allowed.
In this  case the nonlinear term is {\it exactly} zero (see  Fig.
\ref{velocity}b).
Thus as $\nu_x \to \infty$, a slow decrease of $\lambda$ toward zero
is expected.
Our simulations indicate that the decrease
of $\lambda$ is much faster, dropping to immeasurably small values
as we approach the isotropic point, $\nu_x = \nu_y$.

There are  two possible scenarios compatible with
these results. According to the first, $\lambda = 0$ for
$\varepsilon \leq 1$ and  $\lambda \neq 0$ for $\varepsilon > 1$. The
second scenario says that  $\lambda \to 0$ as  $\varepsilon \to 0$.
Our data support the first hypothesis, but at this point we cannot
exclude the second.

{\bf (c)} Figure \ref{velocity}c
shows the results of our simulations when the RSOS condition is applied
to the growth of the string: for a given $i$-th bead, if $h_{i\pm 1} -
h_i > 2$ then we increment $h_i \to h_{i} + 1$ no matter the energy
value of the new configuration.
In contrast with the results of {\bf (a)} and {\bf (b)}, in this case
we find that the parabolas
become steeper as $F \to F_c$,
corresponding to an increase in $\lambda$ as the depinning transition is
approached as in Eq. (\ref{lam-div})  \cite{amaral,amaral2,makse2}.

A typical system to which this study may be relevant is the motion of a
single flux
line in a type-II superconductor, directed along the
external magnetic field $H$. At moderate fields, when the
separation of the vortex lines is sufficiently large, the intervortex
interaction can be neglected. In this regime, the  dynamics of the vortex phase
can be understood by studying the motion of a single vortex.
We argue that
the condition $\nu_x <
\nu_y$ can be  met in some
anisotropic superconductors.
Thus our results might be important in understanding the driven
diffusion of the flux line.
The variation of the
velocity with tilt suggests that a tilted external magnetic field $H$
changes the velocity of the flux line, the effect
decreasing  as we approach the depinning transition.

In summary, we present a model to describe the motion of an elastic
anisotropic
string in a disordered medium. We find that
when transverse fluctuations (along the driving force)
are energetically
more favorable than longitudinal fluctuations,
the string is described by Eq. (\ref{lin}) not only at the
depinning transition but only in the moving phase.
However, if longitudinal fluctuations are more
favorable, a kinematic nonlinear term is induced so that its coefficient
vanishes as we approach the depinning transition.
The directed percolation depinning universality class is obtained when a RSOS
condition is applied that favors the growth of regions with large local
slopes. This last result is shown to be valid for any value of the
anisotropic parameter $\varepsilon$.

We thank L. A. N. Amaral, R. Cuerno,
K. L. Lauritsen, and  S. Tomassone for
valuable discussions. The Center for Polymer Studies is supported by the
NSF.

\begin{figure}
\narrowtext
\caption{The discrete version of the elastic string is composed of $L$
massless beads interacting via springs. A driving external
force $F$ acts in the $y$
direction. Point-like quenched
disorder (not shown) is introduced at each site on the lattice. The
beads are allowed to move in the $x$ and $y$ direction and therefore
they develop overhangs.}
\label{sketch}
\end{figure}

\begin{figure}
\narrowtext
\caption{Plot of average
velocity versus the average tilt of the string for different
values of the reduced force ranging from $f=0.03$ (bottom curves) to
$f=0.20$ (top curves).
Results are averaged over $200$
independent realizations of the disorder. The system size is $L=250$ and
the strength of the disorder is $\delta = 3$.
{\bf (a)} The elastic constants are $\nu_x=0.1$
and $\nu_y=1$ ($\varepsilon = 10$).
The opening of the parabolas as the depinning transition is approached
indicates that a nonlinear
term is present in the equation of motion, and that its value converges to zero
at the depinning transition.
The continuous lines are the best polynomial
fits to the curves.
We note that the observed parabolic dependence cannot be a lattice effect since
in this case
one expect $v(m=0) = v(m=1)$.
{\bf (b)} The same plot for the case when longitudinal fluctuations are not
energetically favorable, so that overhangs are not observed.
The elastic constants
are $\nu_x=1.0$ and $\nu_y=0.1$ ($\varepsilon = 0.1$).
The horizontal lines indicate
that the velocity is independent of the average tilt of the string.
{\bf (c)} Tilt dependence of the average velocity of the string when the RSOS
condition is applied showing the closing of the parabolas indicating a
diverging $\lambda$ term at the depinning transition. Here is shown the
isotropic case $\nu_x = \nu_y=1$ ($\varepsilon = 1$), although
the divergency of $\lambda$ is shown to be independent of the parameter
$\varepsilon$.}
\label{velocity}
\end{figure}



\begin{references}

\bibitem[*]{byline} Present address: University of Notre Dame,
Department of Physics, Notre Dame, IN 46556.


\bibitem{super} G. Blatter, {\it et. al.}  Rev. Mod. Phys. {\bf 66},
1125 (1994); D. Ertas, and M. Kardar, Phys. Rev. Lett. {\bf 73}, 1703
(1994).

\bibitem{Vicsek} A.-L. Barab\'asi and H.~E. Stanley, {\em Fractal
Concepts in Surface Growth\/} (Cambridge University Press, Cambridge,
1995); T. Vicsek, {\it Fractal Growth Phenomena}, 2nd. Edition, (World
Scientific, Singapore, 1992); P. Meakin, Phys.  Rep. {\bf 235}, 189
(1993); T. Halpin-Healy and Y.-C. Zhang, Phys. Rep. {\bf 254}, 215
(1995).


\bibitem{Dong93}
M. Dong, M. C. Marchetti, A. A. Middleton, and V. Vinokur, Phys. Rev.
Lett. {\bf 70}, 662 (1993).

\bibitem{Kaper93}
H. Kaper, G. Leaf, D. Levine, and V. Vinokur,  Phys. Rev.
Lett. {\bf 71}, 3713 (1993).

\bibitem{Leschorn93}
H. Leschorn, and L.-H. Tang, Phys. Rev. Lett. {\bf 70}, 2973 (1993).

\bibitem{lech2}
H. Leschhorn, Physica A {\bf 195}, 324 (1993).

\bibitem{Chang94}
C. Tang, S. Feng and L. Golubovic, Phys. Rev. Lett. {\bf 72}, 1264 (1994).

\bibitem{makse1}
H. A. Makse and L. A. N. Amaral, Europhys. Lett. 379 {\bf 31} (1995).

\bibitem{Natter92}
T. Nattermann, S. Stepanov, L.-H. Tang, and
H.  Leschorn, J. Phys. II
(France) {\bf 2}, 1483 (1992).


\bibitem{Narayan93}
O. Narayan and D. Fisher, Phys. Rev. B {\bf 48}, 7030 (1993).

\bibitem{ew}
S.~F. Edwards and D.~R. Wilkinson, Proc. R. Soc. London Ser.  {\bf A} 381, 17
(1982).

\bibitem{Parisi92}
G. Parisi, Europhys. Lett. {\bf 17}, 673 (1992).

\bibitem{KPZ}
M. Kardar, G. Parisi, and Y.-C. Zhang, Phys. Rev. Lett.
{\bf 56}, 889 (1986).

\bibitem{kk}
J.~M. Kim and J.~M. Kosterlitz, Phys. Rev. Lett. {\bf 62}, 2289
(1989).

\bibitem{lech3}
L.-H. Tang and H. Leschhorn, Phys. Rev. A {\bf 45}, R8309 (1992).

\bibitem{sergey}
S.~V. Buldyrev, A.-L. Barab\'asi, S. Havlin, F. Caserta, H.~E. Stanley
and T. Vicsek, Phys. Rev. A {\bf 45}, R8313 (1992).


\bibitem{snepen}
K. Sneppen, Phys. Rev. Lett. {\bf 69}, 3539 (1992); L.-H. Tang and
H. Leschhorn, {\it Ibid.\/} {\bf 70}, 3832 (1993); K. Sneppen and
M.~H. Jensen, {\it Ibid.\/} {\bf 70}, 3833 (1993).

\bibitem{amaral}
L.~A.~N. Amaral, A.-L. Barab\'asi, and H.~E. Stanley, Phys. Rev. Lett.
{\bf 73}, 62 (1994).

\bibitem{amaral2}
L.~A.~N. Amaral, A.-L. Barab\'asi, H.~A.~Makse, and H.~E. Stanley,
Phys. Rev. E {\bf 52} 4087 (1995).

\bibitem{makse2}
H. A. Makse,
Phys. Rev. E {\bf 52} 4080 (1995).

\bibitem{Krug}
J. Krug and H. Spohn, Phys. Rev. Lett. {\bf 64}, 2332 (1990).

\bibitem{largeslope}
Eq. (\ref{vel}) is expected to be valid for small tilts ($m \sim 0$) only.
For larger tilts higher order nonlinear terms (as $\lambda_4 (\nabla
y)^4$) contribute to  the equation of motion (\ref{kpz}),
and the velocity deviates from the parabola (\ref{vel}). These terms are
not relevant regarding the scaling behavior of the string.



\end{references}
\end{document}